\documentclass[%
 reprint,
showpacs,
 amsmath,amssymb,
 pre,
]{revtex4-1}

\pdfoutput=1
\usepackage{graphicx}
\usepackage{dcolumn}
\usepackage{bm}
\usepackage{wrapfig}
\usepackage{amssymb}
\usepackage{epstopdf}
\usepackage[top=0.75in, bottom=0.75in, left=0.5in, right=0.5in]{geometry}
\usepackage{enumitem}
\setlist[enumerate]{itemsep=0mm}

\begin{document}


\title{A Fokker-Planck approach to graded information propagation\\ in pulse-gated feedforward neuronal networks}
\author{Cong Wang$^{1}$}\email{Co-first authors.}
\author{Zhuocheng Xiao$^{1,*}$}
\author{Zhuo Wang$^1$}
\author{Andrew T. Sornborger$^{2}$}\email{ats@math.ucdavis.edu, taolt@mail.cbi.pku.edu.cn}
\author{Louis Tao$^{1,3,\dagger}$}
\affiliation{$^1$Center for Bioinformatics, National Laboratory of Protein Engineering and Plant Genetic Engineering, College of Life Sciences, Peking University, Beijing, 100871, China}
\affiliation{$^2$Department of Mathematics, University of California, Davis, CA 95616, USA}
\affiliation{$^3$Center for Quantitative Biology, Peking University, Beijing, 100871, China}

\date{\today}

\begin{abstract}
\noindent 
Information transmission is a key element for information processing in the brain. A number of mechanisms have been proposed for transferring volleys of spikes between layers of a feedforward neural circuit. Many of these mechanisms use synchronous activity to provide windows in time when spikes may be transferred more easily from layer to layer. Recently, we have demonstrated that a pulse-gating mechanism can transfer graded information between layers in a feedforward neuronal network. Our transfer mechanism resulted in a time-translationally invariant firing rate and synaptic current waveforms of arbitrary amplitude, thus providing exact, graded information transfer between layers. In this paper, we apply a Fokker-Planck approach to understand how this translational invariance manifests itself in a high-dimensional, non-linear feedforward integrate-and-fire network. We show that there is good correspondence in spiking probabilities between the Fokker-Planck solutions and our previous mean-field solutions. We identify an approximate line attractor in state space as the essential structure underlying the time-translational invariance of our solutions. This approach has enabled us to better understand the role played by the synaptic coupling, gating pulse and the membrane potential probability density function in information transfer.
\end{abstract}


\maketitle


\section{Introduction}

Time-translationally-invariant spiking probabilities are signatures of information propagation in neural systems. Coherence is a measure of information transmission and coherent spiking activity has been measured experimentally in many regions of the brain \citep{GrayEtAl1989,Livingstone1996,WomelsdorfEtAl2007,BroschEtAl2002,BauerEtAl2006,PesaranEtAl2002,BuschmanMiller2007,MedendorpEtAl2007,BuschmanMiller2007,GregorgiouEtAl2009,SohalEtAl2009,OKeefe1993,Buzsaki2002}. Many theoretical and experimental studies have shown that synchronized volleys of spikes can propagate within cortical networks and are thus capable of transmitting information between neuronal populations on millisecond timescales \cite{pmid12684488,pmid10591212,pmid12730700,KistlerGerstner2002,pmid24298251,pmid16641232}. Mechanisms that have been proposed for information transfer include synfire chains \cite{pmid10591212,pmid12730700,KistlerGerstner2002}, networks operating at firing threshold \cite{pmid11880526}, and networks with oscillatory input at resonant frequencies \cite{pmid24730779,pmid25503492,AkamKullmann2010,pmid24434912}. The goal of these modeling studies has largely been to explain experimental results \cite{pmid11244543,pmid22007180} demonstrating that information can be rapidly cascaded through multilayer, feedforward networks in the brain \cite{pmid12684488,pmid10591212,KistlerGerstner2002}.

The precise mechanism and the extent to which the brain can make use of synchronous spiking activity to transfer information have remained unclear. Recently, we put forward a pulse-gating mechanism for information propagation in neural circuits that consists of a pulse-generating component and an information-containing component \cite{SornborgerWangTao,WangSornborgerTao}. Graded information, in packets of spikes, is propagated from an upstream population to a downstream population when a pulse brings the upstream population to threshold, causing it to fire, and downstream neuronal populations to integrate the incoming spikes. In the mean, this mechanism gives rise to time-translationally-invariant spiking probabilities that are exactly propagated from layer to layer.

The original information propagation model that we put forward was based on a mean-field firing rate model of a network of current-based integrate-and-fire (I\&F) neurons. We demonstrated that there was good correspondence between the mean-field model and a network of I\&F neurons. However, it remained to be understood how the mean-field model gave such good correspondence with the I\&F model. In this paper, in order to better understand this relationship, we have constructed Fokker-Planck equations describing the membrane potential's probability density function in the pulse-gating context and studied their solutions.

In Section \ref{sec2}, we describe the I\&F model and the resulting Fokker-Planck equations. In Section \ref{sec3}, we simulate the Fokker-Planck equations and analyze the state-space of graded information propagation. Finally, in Section \ref{sec4}, we discuss our results and their implications for information coding.

\section{Model and Fokker-Planck Description}\label{sec2}

We use a neuronal network model consisting of a set of $j = 1, \dots, M$ populations, each with $i = 1, \dots, N$ excitatory, current-based, integrate-and-fire (I\&F) point neurons whose membrane potential and (feedforward) synaptic current are described by
\begin{widetext}
\begin{subequations}
\begin{eqnarray}
 \frac{d}{{dt}}V_{i,j}^{} & = &  - {g_L}\left( {V_{i,j}^{} - {V_R}} \right) + I_{j}^g + I_{i,j}^{ff} \label{IFa}\\ 
 \tau \frac{d}{{dt}}I_{i,j}^{ff} & = &  - I_{i,j}^{ff} + \left\{ \begin{array}{ll} \frac{S}{{pN}}\sum\limits_k^{} p_{jk}{\sum\limits_l {\delta \left( {t - t_{j-1,k}^l} \right)}}, & j = 2, \dots, M \\ A \delta(t), & j = 1 \end{array} \right. \label{IFb}
\end{eqnarray}
\end{subequations}
\end{widetext}
where $V_R$ is the reset voltage, $\tau$ is the synaptic timescale, $S$ is the synaptic coupling strength, $p_{jk}$ is a Bernoulli distributed random variable and $p = \langle p_{jk} \rangle_{jk}$ is the mean synaptic coupling probability. The $l$'th spike time of the $k$'th neuron in layer $j-1$ is determined by $V(t^l_{j-1,k}) = V_{Th}$, i.e. when the neuron reaches threshold. The gating current, $I^g_{j}$, is a white noise process with a square pulse envelope, $\theta(t - (j-1)T) - \theta(t - jT)$, where $\theta$ is a Heaviside theta function and $T$ is the pulse length \cite{SornborgerWangTao} of pulse height $\bar{I}^g$ and variance $\sigma_0^2$. Note that with the $j = 1$ equation, an exponentially decaying current is injected in population $1$ providing graded synchronized activity that will subsequently propagate downstream through populations $j = 2, \dots, M$.

\begin{figure}[t]
  \includegraphics[width=0.5\textwidth]{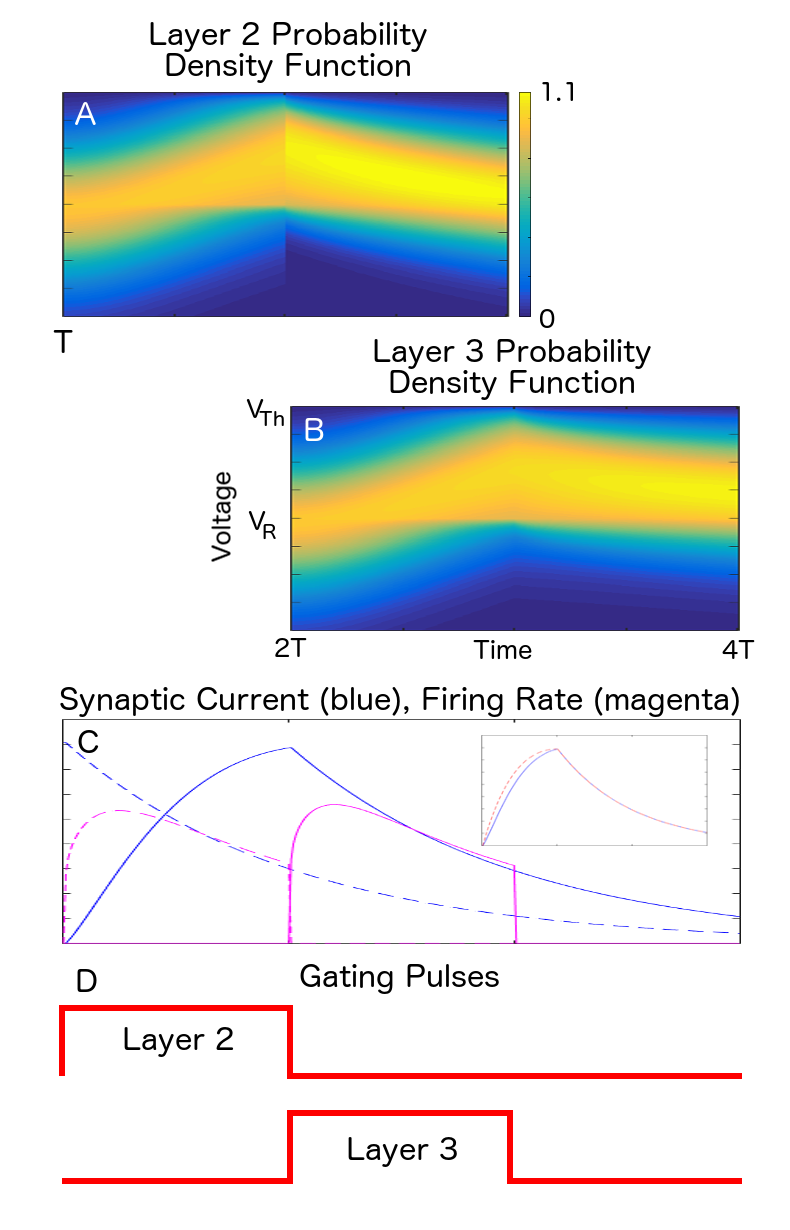}
  \caption{Fokker-Planck dynamics. Evolution of $\rho_j(V,t)$ for the second and third layer in a $12$ layer simulation. For each gating pulse, $T = 5$ ms. $S = 2.9$, $\sigma_0^2 = 20$, $\bar{I}^g = 13.5$. A \& B) $\rho_j(V,t)$, $j = 2,3$ resp. with $t = (j-1)T$ to $jT$. C) Firing rates (magenta), $m_j$, and synaptic currents (blue), $\bar{I}^{ff}_j$, for layers $2$ (dashed) and $3$ (solid). Inset: Comparison between mean-field (dashed red) and Fokker-Planck (blue) feedforward synaptic currents. D) The gating pulses for layers $2$ and $3$. Offset vertically for clarity.}
\end{figure}

Assuming the spike trains in Eq. (\ref{IFb}) to be Poisson distributed, the collective behavior of this feedforward network may be described by the Fokker-Planck equations
\begin{subequations}
\begin{eqnarray}
 \frac{\partial}{{\partial t}}{\rho _j}\left( {V,t} \right) & = & -\frac{\partial}{\partial V}J_j(V,t)\label{FP1} \\ 
 \tau \frac{d}{{dt}}\overline I _j^{ff} & = & -\overline I _j^{ff} + \left\{ \begin{array}{ll} S{m_{j - 1}}, & j = 2, \dots, M \\ A\delta(t), & j = 1 \end{array} \right. \label{FP2}
\end{eqnarray}
\end{subequations}
These equations describe the evolution of the probability density function, $\rho_j(V,t)$, in terms of the probability density flux, $J_j(V,t)$, the mean feedforward synaptic current, $\bar{I}^{ff}_j$, and the population firing rate, $m_j$. For each layer, $j$, the probability density function gives the probability of finding a neuron with membrane potential $V \in (-\infty,V_{Th}]$ at time $t$.

The probability density flux is given by
\begin{eqnarray}
J_j\left( {V,t} \right) & = & 
 \left( \left[ { - {g_L}\left( {V - {V_R}} \right) + \overline I ^g + \overline I _j^{ff}} \right] \right. \;\;\;\;\;\;\; \nonumber \\
 & & - \left. {\sigma _j^2}\frac{\partial }{{\partial V}} \right){\rho _j}\left( {V,t} \right)\;,
\end{eqnarray}
where $\bar{I}^g$ indicates the mean gating current. The effective diffusivity is
\begin{equation}
\sigma _j^2 = \sigma_0^2 + \frac{1}{2}{\frac{{{S^2}}}{{pN}}{m_{j - 1}}\left( t \right)}\;.
\end{equation}
(In the simulations reported below, we take $N \rightarrow \infty$.) The population firing rate is the flux of the probability density function at threshold,
\begin{equation}
{m_j}\left( t \right) = J_j\left( {{V_{Th}},t} \right)\;.
\end{equation}

The boundary conditions for the Fokker-Planck equations are
\begin{eqnarray}
J_j\left( {V_R^ + ,t} \right) & = & J_j\left( {{V_{Th}},t} \right) + J_j\left( {V_R^ - ,t} \right)\;,\\
{\rho _j}\left( {V_R^ + ,t} \right) & = & {\rho _j}\left( {{V_{Th}},t} \right) + {\rho _j}\left( {V_R^ - ,t} \right)\;,
\end{eqnarray}
and
\begin{equation}
{\rho _j}\left( {V =  - \infty ,t} \right) = 0\;.
\end{equation}

\section{Results}\label{sec3}

We simulated the evolution of the probability density function across many layers. Figure 1 shows the dynamics of the probability density function, $\rho_j$, the firing rate, $m_j$, and the feedforward synaptic current, $\bar{I}^{ff}_j$, in layers $j = 2$ and $3$ of the simulated network. As is shown in Fig. 1A, the combined effect of the gating (Fig. 1D) and the feedforward synaptic current in layer $2$ (Fig. 1C, dashed blue) drives $\rho_2$ toward threshold, leading to spiking activity (Fig. 1C, dashed magenta) and inducing a synaptic current in layer $3$ (Fig. 1C, solid blue). Termination of gating at $t = 2T$ leads to a decay in $\rho_2$ towards $V_R$, abolishing firing. Consequently, the synaptic current of layer $3$ decays exponentially. The combination of this exponentially decaying synaptic current and the subsequent gating pulse causes firing in layer $3$ (Fig. 1C, solid magenta), driving downstream layers.

We compared synaptic currents for the mean-field \cite{SornborgerWangTao} and Fokker-Planck equations (Fig. 1C, inset). The onset of the rise in synaptic current in the Fokker-Planck solution was slower than the mean-field solution. This was due to the non-zero time to reach threshold of the probability density function, $\rho_j(V,t)$, in the Fokker-Planck solution. Other than this small difference, both synaptic currents decayed exponentially once gating was terminated, and the solutions were markedly similar.

\begin{figure}[t]
  \includegraphics[width=0.5\textwidth]{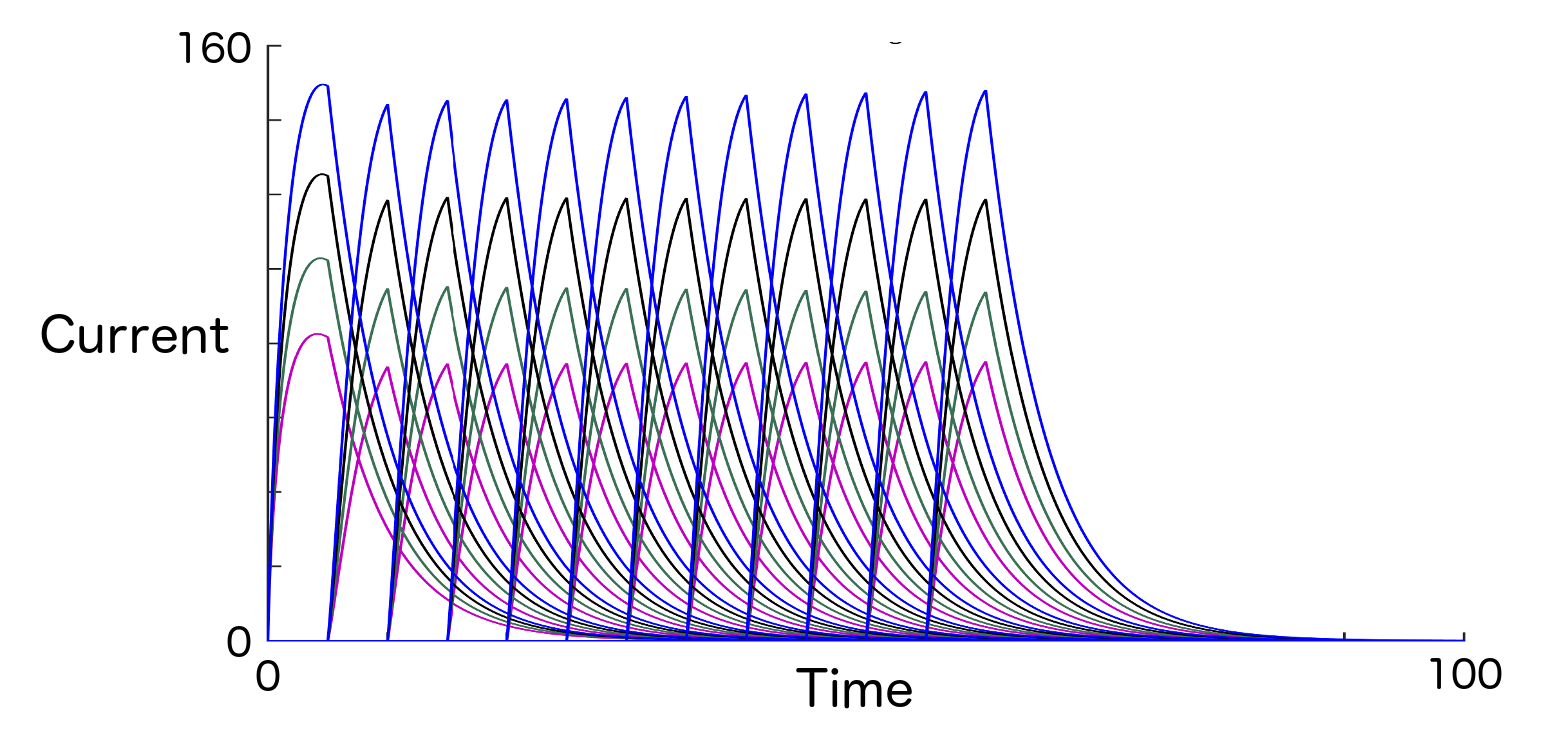}
    \caption{Propagation of graded activity across $12$ layers. $S = 2.9$, $\sigma_0^2 = 20$, $\bar{I}^g = 13$. Here, we show the induced current, $\bar{I}^{ff}$ for $j = 1,\dots,12$, on downstream layers due to upstream firing, $m_{j-1}(t)$.}
\end{figure}

\begin{figure*}[t]
  \includegraphics[width=\textwidth]{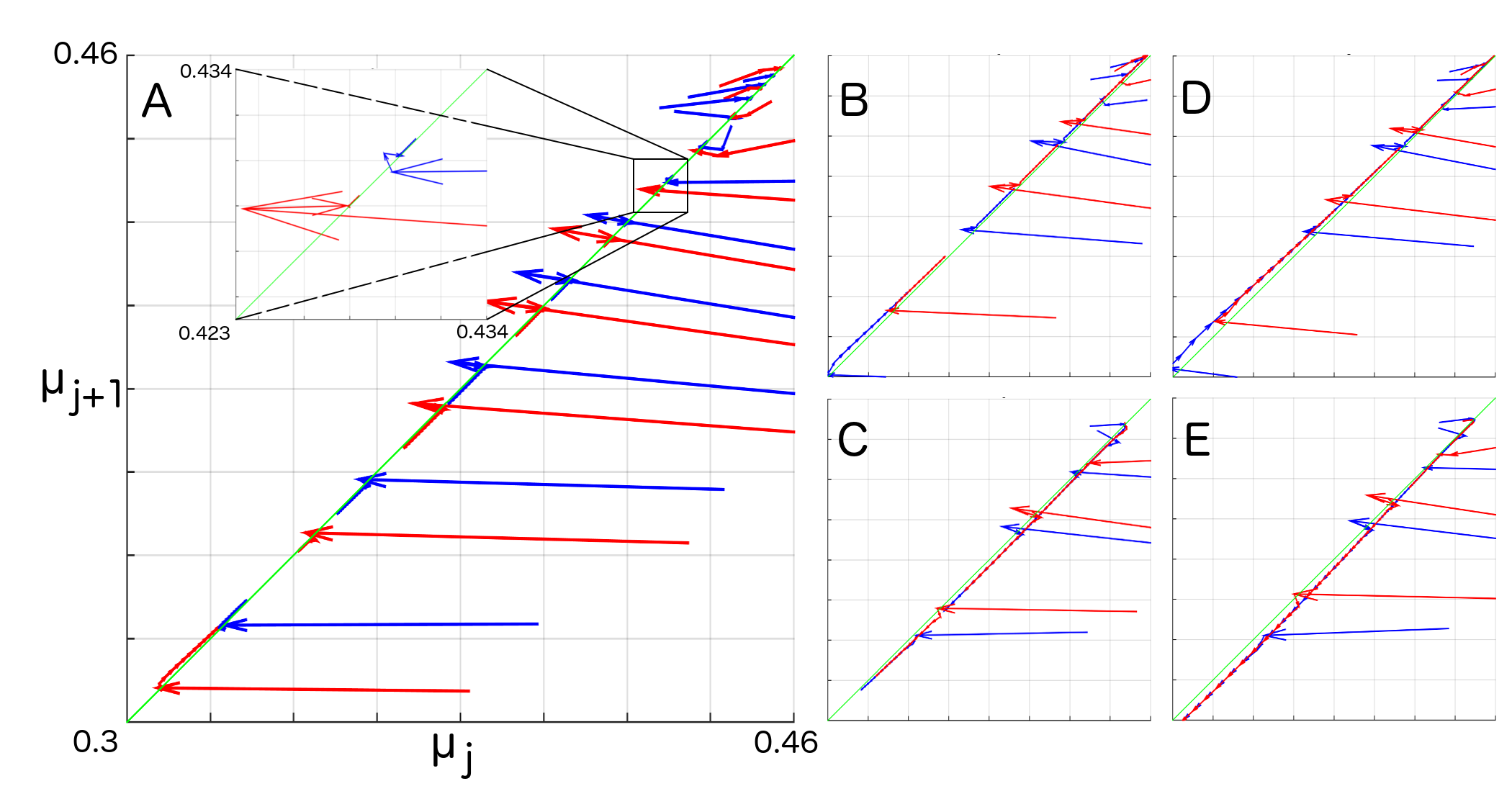}
   \caption{State-space analysis of graded propagation. $\mu_{j}$ is plotted versus $\mu_{j+1}$ with trajectories for given initial conditions shown in red or blue arrows. $\sigma_0^2 = 20$ in all panels. A) Graded propagation: $S = 2.9$, $\bar{I}^g = 13$. Here, for most initial conditions, trajectories are strongly attracted to the diagonal. After an initial transient, $\mu$ is approximately fixed (See Fig. 2). The inset shows the slow dynamics of $\mu$ for two amplitudes. Non-graded propagation: B) $S = 3.0$, $\bar{I}^g = 13$. C) $S = 2.8$, $\bar{I}^g = 13$. D) $S = 2.9$, $\bar{I}^g = 16$. E) $S = 2.9$, $\bar{I}^g = 10$.}
\end{figure*}

To investigate whether graded propagation was possible in the Fokker-Planck setting, we searched for values of $S$, $\bar{I}^g$ and $\rho_j(V,0)$ that allowed a range of amplitudes to be propagated without change through many layers. In Fig. 2, we examine an instance of graded propagation. We show synaptic currents corresponding to four initial amplitudes propagating across $12$ layers. After an initial transient, the synaptic current waveforms are stereotypical and the maximum fractional change in amplitude (which occurs for the largest amplitude depicted) is approximately $0.002$ (i.e. $0.2 \%$) per transfer. Firing rate waveforms (not shown here, but see Fig. 1C) are also stereotypical with non-zero time to threshold and sharp termination due to pulse gating. Because of the slow onset of firing, the value of synaptic coupling that gave rise to graded transfer, $S = S_{graded}$, was larger than the mean-field prediction, $S_{exact}$ (see Appendix), by a factor of $1.07$.

The discrete, time-translational invariance of the synaptic currents and firing rates when graded information propagates through a feedforward network rests on the time-translational invariance of the underlying probability density function, {\it viz}.
\begin{equation}
  \rho_j(V,t - (j-1)T) = \rho_{j+1}(V,t-jT) \; .
\end{equation}
Notice in Fig. 1A,B that $\rho_3(V,t-2T)$ is very similar to $\rho_2(V,t-T)$. In fact, for exact translational invariance, the dynamics of $\rho_j$ become stereotypical, and may be entirely characterized by the probability density function at the termination of gating $\rho_j(V,jT)$ and the probability density flux at threshold during gating, i.e. $m_j(t)$ for $t \in [(j-1)T,jT]$.

The probability distribution function itself is high dimensional, however, its main features can be captured by examining its first few moments. In Fig. 3 (all panels), we plot the mean of the distribution at the end of pulse-gating, 
\begin{eqnarray}
\mu_j & = & \mathbb{E}\left\{ V; \rho_j(V,jT) \right\} \nonumber \\
& = & \int_{-\infty}^{V_T} V \rho_j(V, jT) dV \; ,
\end{eqnarray}
of successive layers against each other. For exact, time-translational invariance, $\mu_j = \mu_{j+1}$, and trajectories in state-space, $( \mu_j, \mu_{j+1})$, would consist of fixed points on the diagonal.

In Fig. 3A, we show trajectories in state-space for an instance of graded propagation. For all initial conditions depicted, trajectories rapidly approach, then slowly drift along the diagonal (see Fig. 3A, inset). The slow changes in $\rho_j$ that give rise to the drift of $\mu_j$ along the diagonal induce the small drifts in the amplitudes of synaptic currents seen in Fig. 2. These dynamics indicate the existence of a saddle node with unstable manifold along the diagonal. The system rapidly converges to the unstable manifold and the slow dynamics along this manifold give rise to an approximate line attractor in this space. In Fig. 3B-E, we investigate the behavior of the system when graded propagation fails. Roughly, for $S > S_{graded}$, $\mu_j$ increases over time (Fig. 3B), and for $S < S_{graded}$, $\mu_j$ decreases over time (Fig. 3C). We note that the dynamics are also dependent on $\bar{I}^g$. So, varying around $\bar{I}^g = \bar{I}_{graded}$, we see that, for $\bar{I}^g > \bar{I}^g_{graded}$, $\mu_j$ increases over time (Fig. 3D), and for $\bar{I}^g < \bar{I}^g_{graded}$, $\mu_j$ decreases over time (Fig. 3E). From these results, mechanisms of failure are apparent, either the unstable manifold is no longer along the diagonal, or the dynamics on the unstable manifold are no longer slow.

\section{Discussion}\label{sec4}

Using the tools of Fokker-Planck analysis for stochastic systems, we have investigated the correspondence between graded information transfer in an I\&F neuronal network and its mean-field firing rate model. We have demonstrated that there is a close correspondence between solutions of both models. We have also pointed out small differences between the models and the reasons for these differences.

Inherently, the dynamics of the I\&F network is high-dimensional. However, the Fokker-Planck analysis gives a clear understanding of why the mean-field model works so well. It is because, for parameters supporting graded transfer, 1) the synaptic currents and firing rates approach fixed waveforms, and 2) the underlying probability density functions also admit an approximate discrete translational invariance after an initial transient.

We gave evidence for the translational invariance of the probability density function via a state-space analysis of its mean in successive layers. Using this, we found that graded propagation led to rapid convergence to the diagonal, i.e. an approximate line attractor in this state space. Line attractors have been appealed to in attractor network models where patterns of spiking activity lie on a line of fixed points in some state space \cite{pmid10798409,pmid8917592,pmid15718474} and they have been observed experimentally \cite{pmid24201281}. Typically, line attractors arise in contexts where neuronal encoding requires some sort of gradedness in a one-dimensional variable. This is also true of our system, and here we have been able to show how a line attractor can manifest itself in an integrate-and-fire network and studied its dependences on a low-dimensional state space identified by the Fokker-Planck approach.

%
Our previous mean-field analysis contained only one parameter on which graded propagation relies, the synaptic coupling, $S_{exact}$. However, from the Fokker-Planck analysis we now understand how, in addition to the synaptic coupling, $S$, the gating current, $\bar{I}^g$, and the initial probability density function, $\rho_j(V,jT)$, also play an essential role. This implies that there may be a range of parameters for which graded propagation is possible making pulse-gated information propagation more robust. It appears that, for a wide range of $S$ near $S_{exact}$, there will be an $\bar{I}^g$ and $\rho_j(V,jT)$ that will give rise to graded propagation. If so, the regulation and control of neural pathways supporting graded information transmission could make use of spike-timing dependent plasticity (to regulate $S$ and $\bar{I}^g$), network coherence (to regulate $\bar{I}^g$), and noise backgrounds (to regulate $\rho_j(V,jT)$).

\section{Appendix: Mean Field Model}

A mean-field firing rate model of Eq. (1) is given by
\begin{eqnarray}
  \tau \frac{d\bar{I}^{ff}_j}{dt} & = & -\bar{I}^{ff}_j + Sm_{j-1} \\
  m_j & = & \left[ \bar{I}^{ff}_j + I^g_j(t) - g_0 \right]^+
\end{eqnarray}
where we use a thresholded linear approximation, with threshold $g_0$, for the input-output relation of an I\&F neuron.

In \cite{SornborgerWangTao}, we showed that when the gating pulse cancels the threshold ($\bar{I}^g = g_0$), and the feedforward synaptic coupling strength was
\begin{equation}
  S_{exact} = \frac{\tau}{T}e^{T/\tau}
\end{equation}
we get exact, graded propagation, where the mean synaptic current and firing rates were
\begin{widetext}
\begin{equation}
  I^{ff}_j(t) = \left\{ \begin{array}{ll}
           A\left( \frac{t - (j - 1)T}{\tau} \right) e^{-(t - (j-1)T)/\tau}, & (j-1)T \le t \le jT \\
           A \left( \frac{T}{\tau} e^{-T/\tau} \right) e^{-(t - jT)/\tau}, & jT < t < \infty \end{array} \right.
\end{equation}
and
\begin{equation}
  m_j(t) = \left\{ \begin{array}{ll}
           0, & 0 < t <(j-1)T \\
           A \left( \frac{T}{\tau} e^{-T/\tau} \right) e^{-(t - jT)/\tau}, & (j-1)T \le t \le jT \\
           0, & jT < t < \infty \end{array} \right.
\end{equation}
\end{widetext}

\begin{acknowledgments}
L.T. thanks the UC Davis Mathematics Department for its hospitality. A.T.S. would like to thank Liping Wei and the Center for Bioinformatics at the College of Life Sciences at Peking University
for their hospitality. This work was supported by the Ministry of Science and Technology of China through the Basic Research Program (973) 2011CB809105 (C.W., Z.X., Z.W. and L.T.), by the Natural Science Foundation of China grant 91232715 (C.W., Z.X., Z.W. and L.T.) and by the National Institutes of Health, CRCNS program NS090645 (A.T.S. and L.T.).
\end{acknowledgments}

\bibliographystyle{unsrtnat}
\bibliography{Biblio}

\end{document}